\documentclass[epj,twocolumn]{webofc}
\usepackage[varg]{txfonts}   
\woctitle{Hadron Collider Physics symposium 2012}
\begin{document}
\title{SUSY vs LHC}
%
%

\author{Ryuichiro Kitano\inst{1}\fnsep\thanks{\email{kitano@tuhep.phys.tohoku.ac.jp}} 
}

\institute{Department of Physics, Tohoku University, Sendai 980-8578, Japan
          }

\abstract{ In light of the discovery of the new particle at 125~GeV and
the strong lower limits on the masses of superparticles from LHC, we
discuss a possible picture of weak scale supersymmetry.  }
\maketitle
\section{Introduction}
\label{intro} The searches for superparticles at the LHC have been
putting severe constraints on weak scale supersymmetry. The lower bound
on the gluino mass from searches for jets with missing energy is
roughly~\cite{atlas-gluino, cms:2012mfa},
\begin{eqnarray}
 m_{\tilde g} \gtrsim {1-1.5}~{\rm TeV}.
\end{eqnarray}
A similar constraint applies for squarks in the first generation. The
constraints on the masses of the scalar top and bottom quarks (stop and
sbottom) are slightly weaker such as
\begin{eqnarray}
 m_{\tilde t,\tilde b} \gtrsim 500-600~{\rm GeV},
\end{eqnarray}
for a non-degenerate neutralino~\cite{atlas-stop, cms-stop}.

The new particle at 125~GeV~\cite{atlas:2012gk, cms:2012gu} is also
giving an important constraint on SUSY models. If the particle is
interpreted as the lightest Higgs boson in the minimal supersymmetric
standard model (MSSM), the stop mass is required to be
\begin{eqnarray}
 m_{\tilde t} \gtrsim {\mbox{(a few)}} \times~{\rm TeV}.
\label{eq:stop}
\end{eqnarray}

In light of these constraints, we discuss what kind of theoretical
framework is implied by the LHC data if SUSY is the solution to the
naturalness problem.

\section{The log problem}
\label{sec-1}

In softly broken SUSY models, the quadratic divergence is canceled, and
that is supposed to be the solution to the hierarchy problem.
As one can see below, that is not quite enough any more. We are facing
the situation where one also needs to eliminate the log divergence.

There is a logarithmic divergence in the quantum corrections to mass
parameters in supersymmetric standard model. Once we put a large gluino
mass, $m_{\tilde g}$, at a high scale $M$, the stop mass squared and the
quadratic term in the Higgs potential (which is the Higgs mass squared
times ($-1/2$)), receives quantum corrections proportional to $m_{\tilde
g}^2 \log (M/m_{\tilde g})$. If the log factor is large and if there is
no significant cancellation among parameters, we naively expect
\begin{eqnarray}
 m_h^2 \sim m_{\tilde t}^2 \sim m_{\tilde g}^2.
\label{eq:guess}
\end{eqnarray}
This is clearly inconsistent with $m_h \sim 125$~GeV and $m_{\tilde g}
\gtrsim 1$~TeV. This is the SUSY fine-tuning problem.
Especially, in models which are friendly with the grand unification,
such as gravity mediation, the log factor tends to be large.
In the MSSM where the Higgs mass is obtained by the gauge coupling times
the VEV of the Higgs field at tree level, the above relation either
predicts an unacceptably large Higgs VEV or light gluino/stop that is no
longer allowed by the LHC data.


In order to correct the wrong prediction in Eq.~\eqref{eq:guess}, the
logarithmic quantum correction should be cut-off at a scale close to the
TeV scale.
In that case, the size of the quantum corrections are estimated to be
\begin{eqnarray}
 \delta m_h^2 \sim {y_t^2 N_c \over 8 \pi^2} m_{\tilde t}^2 \sim (0.2)^2
  m_{\tilde t}^2
\end{eqnarray}
\begin{eqnarray}
 \delta m_{\tilde t}^2 \sim {g_3^2 \over 4 \pi^2}{N_c^2 - 1 \over N_c}
 m_{\tilde g}^2 \sim (0.3)^2 m_{\tilde g}^2.
\end{eqnarray}
Therefore, a mild hierarchy such as
\begin{eqnarray}
 m_h \sim 0.2 m_{\tilde t},\ \ \  m_{\tilde t} \sim 0.3 m_{\tilde g},
\end{eqnarray}
is possible without fine-tuning. For example, 
\begin{eqnarray}
 m_{\tilde g} \sim 2~{\rm TeV}, \ \ \ m_{\tilde t} \sim 600~{\rm GeV}, \
  \ \ m_h \sim 120~{\rm GeV},
\label{eq:susy}
\end{eqnarray}
is realized naturally. The spectrum is consistent with the Higgs mass
and barely satisfies the LHC searches for superparticles.
At this stage, one needs to give up the MSSM since the spectrum is
inconsistent with Eq.~\eqref{eq:stop}.

Therefore, in light of various news from LHC, a theoretical framework
to realize natural SUSY requires:
\begin{itemize}
 \item the cut-off of the quantum correction at TeV, and
 \item a new contribution to the Higgs boson mass beyond the stop loop.
\end{itemize}
Both of the requirements actually say that we should have a new physics
beyond the MSSM at the TeV energy scale.

\section{Naive Dimensional Analysis and Partially Composite Higgs}

One of the most drastic ideas for the TeV new physics beyond the MSSM is
the scenario where the Higgs boson is (partially) a composite particle.
A TeV scale strong dynamics is assumed to provide an MSSM like theory as
the low energy effective theory. Because of the drastic change of the
description at the TeV scale, one may expect a cut-off of the log
divergence and a new contribution to the Higgs mass, simultaneously.

In fact, modern efforts of SUSY model building have started with this
type of models~\cite{Witten:1981nf,Dine:1981za,Dimopoulos:1981au}. SUSY
is introduced to protect the quadratic divergence of the Higgs potential
whereas the electroweak scale is generated by some dynamics at TeV as in
the technicolor model. Such models are now well motivated to be
considered again given that there is a light Higgs boson which is not
quite as light as the MSSM predictions.

Once we assume that there is some dynamics at a scale $\Lambda$
responsible for the electroweak symmetry breaking, the naive dimensional
analysis~\cite{Manohar:1983md,Luty:1997fk} says
\begin{eqnarray}
 \Lambda \sim 4 \pi v \sim 3~{\rm TeV},\ \ \ m_h \sim \Lambda.
\end{eqnarray}
where $v = 246$~GeV. This is the typical prediction of the technicolor
models. Now, for having a light Higgs boson, one can assume that the
Higgs fields are slightly weakly coupled to the dynamics at the scale
$\Lambda$. By introducing a dimensionless parameter $\epsilon$ which
measures the weakness of the coupling compared to the naive estimates,
we have
\begin{eqnarray}
 \Lambda \sim 4 \pi \epsilon v,\ \ \ m_h \sim \epsilon \Lambda \sim
  m_{H},
\label{eq:nda}
\end{eqnarray}
where $m_{H}$ is the size of the soft SUSY breaking contribution
to the Higgs potential. In the MSSM language, $\epsilon \Lambda$ is the
$\mu$-term and $m_{H}$ is the soft SUSY breaking mass.
From Eq.~\eqref{eq:nda}, we have
\begin{eqnarray}
 {m_h \over v} \sim 4 \pi \epsilon^2.
\end{eqnarray}
By putting $m_h = 125$~GeV, we obtain $\epsilon \sim 0.2$ and $\Lambda
\sim 600$~GeV.
If the top quark is also involved in the dynamics, the estimate of the
dynamically generated top quark mass is
\begin{eqnarray}
 m_t \sim \epsilon_t^2 \Lambda,
\end{eqnarray}
where $\epsilon_t$ again measures the weakness of the coupling to the
strong sector. For $m_t \sim 170$~GeV, we have
\begin{eqnarray}
 \epsilon_t \sim 0.5.
\end{eqnarray}

From these estimates, we have a rough picture:
\begin{itemize}
 \item there is some dynamics at $\Lambda \sim 600$~GeV,
 \item the Higgs fields (and possibly the top quark) are weakly coupled
       to the dynamics with a suppression factor $\epsilon \sim 0.2$
       (and $\epsilon_t \sim 0.5$), and 
 \item above the scale $\Lambda$, the picture drastically changes so
       that the log divergences of the soft SUSY breaking parameters are
       cut-off.
\end{itemize}
The Higgs quartic coupling and the Higgsino mass are both dominated by the
dynamically generated potential. This is very different from the MSSM
case where the Higgs potential is mainly from the $SU(2)_L \times
U(1)_Y$ gauge interactions ($D$-term) and the $\mu$-term is added by
hand.

The parameters $\epsilon$ and $\epsilon_t$ can be thought of as the
degree of compositeness. The relations among $v$, $m_h$, and $m_t$
suggest that the Higgs/top sector is not fully composite, but maybe
partially composite~\cite{Kitano:2012wv}. The partially composite
scenario is realized only when there is some reason for the absence of
the tree-level Higgs potential of $O(\Lambda)$. SUSY is giving a good
reason for it since one can naturally have an approximately flat
potential even at the quantum level. The (approximately) massless
composite particle is ubiquitous in SUSY gauge theories.

\section{A model for dynamical electroweak symmetry breaking}
\label{sec:model}

A pretty simple model of partially composite Higgs can be
constructed~\cite{Fukushima:2010pm, Kitano:2012cz}.
The dynamical sector is an ${\cal N}=1$ SUSY $U(2)$ gauge theory with
five flavors of chiral superfields.
The assignment of the quantum numbers of the standard model gauge group,
$(SU(3),SU(2))_Y$, for the matter fields is
\begin{eqnarray}
 f_A : (1,2)_{0},\ \ \ \bar f_A : (1,1)_{\pm 1/2},\ \ \ (A=1,2),
\end{eqnarray}
\begin{eqnarray}
 f_\alpha^\prime : (3,1)_{1/6},\ \ \ \bar f_\alpha^\prime : (\bar 3,1)_{-1/6}, \ \ \
  (\alpha = 3,4,5).
\end{eqnarray}
This theory is in the conformal window~\cite{Seiberg:1994pq}. The Higgs
field, $H = (H_u, H_d)$, and the top quark, $q, t^c$, can couple to this
CFT through the superpotential terms:
\begin{eqnarray}
 W = \lambda_H \bar f H f 
+ \lambda_q \bar f^\prime q f + \lambda_t \bar f t^c f^\prime.
\end{eqnarray}
The gauge coupling $g$ and the coupling constants $\lambda_H$,
$\lambda_q$, and $\lambda_t$ flow to IR fixed points whose values are
estimated by the $a$-maximization~\cite{Intriligator:2003jj} such as:
\begin{eqnarray}
{g \over 4 \pi} \sim 0.4,\ \ \  {\lambda_{H_d} \over 4 \pi} \sim 0.1,\ \ \ 
{\lambda_{H_u} \over 4 \pi} \sim 0.3,
\end{eqnarray}
\begin{eqnarray}
 {\lambda_q \over 4 \pi} \sim 0.3,\ \ \  {\lambda_t \over 4 \pi} \sim 0.3.
\end{eqnarray}
By adding a mass to $f^\prime$ and $\bar f^\prime$ by superpotential:
\begin{eqnarray}
 W = \Lambda \bar f^\prime f^\prime,
\end{eqnarray}
the $SU(2)$ gauge interaction becomes strong and confines at the scale
$\Lambda$. The coupling constants $\lambda$'s receive multiplicative
renormalization, $\lambda \to \lambda k$ with $k>1$.
By turning on SUSY breaking terms by
\begin{eqnarray}
 \Lambda \to \Lambda ( 1 + m_{\rm SUSY}^2 \theta^2),
\end{eqnarray}
the Higgs potential are generated. For $\Lambda \sim m_{\rm SUSY}$, the
lightest Higgs field (which is mainly $H_d$) obtain a mass:
\begin{eqnarray}
 m_h^2  \sim \left({\lambda_{H_d} \over 4 \pi} \right)^2 \Lambda^2
\sim m_h^2|_{\rm MSSM} + {\lambda_{H_d}^4 v^2 \over (4 \pi)^2},
\end{eqnarray}
and the Higgsino mass is also generated as $m_{\tilde H} \sim
(\lambda_{H_u} \lambda_{H_d} / (4 \pi)^2) \Lambda$.
The dynamically generated top quark mass is
\begin{eqnarray}
 m_t \sim {\lambda_q \lambda_t \lambda_{H_d} \over (4 \pi)^2} v.
\end{eqnarray}
The correspondence to discussion in the previous section is
\begin{eqnarray}
 \epsilon \sim {\lambda_{H_d} \over 4 \pi}, \ \ \ 
 \epsilon_t^2 \sim {\lambda_q \lambda_t \over (4 \pi)^2}.
\end{eqnarray}
We see that the required value $\epsilon \sim 0.2$ and $\epsilon_t^2
\sim (0.5)^2$ is roughly consistent with the fixed point values with the
$k$ factor $\sim 2$.

This model should be thought of as the effective theory around the scale
$\Lambda$. In order to cut-off the log divergence proportional to the
gluino mass squared, the semi-weakly coupled description above should be
replaced by a more strongly coupled one. For example, one can assume
that the gauge symmetry is extended to $SU(3)$ above the scale $\Lambda$,
which makes the theory a strongly coupled CFT. In this case, the stop mass
parameter is also strongly renormalized such as
\begin{eqnarray}
 {d \over d \log \mu} m_{\tilde t}^2 
= c m_{\tilde t}^2 
- {g_3^2 \over (4 \pi)^2}{N_c^2 - 1 \over N_c} m_{\tilde g}^2
+ \cdots,
\end{eqnarray}
where $c$ is of $O(1)$. The quasi IR fixed point of the stop mass is
then roughly
\begin{eqnarray}
 m_{\tilde t}^2 \sim 
{g_3^2 \over (4 \pi)^2 c}{N_c^2 - 1 \over N_c} m_{\tilde g}^2,
\end{eqnarray}
which is equivalent to estimate without the log divergence. The same is
true for the effect of the stop mass to the Higgs mass parameters.
One can consider a more drastic change of the theory such as the
appearance of an extra dimension at a scale $\Lambda$. See
Ref.~\cite{Kitano:2012cz} for a proposal to obtain composite Higgs
fields couple to the $U(2)$ model above from a higher dimensional QCD.

\section{Why SUSY?}

We see that there is a simple solution to the SUSY fine-tuning problem
once we cut-off the theory at the TeV scale. But once we give up the
SUSY desert, why do we need SUSY?

Aside from the theoretical beauty or calculability, we should remember
that the presence of the light Higgs boson already provides a good
motivation.
The naive dimensional analysis says that relation between the decay
constant $v$ and the mass of the $0^+$ resonance, $m_h$, is estimated to
be
\begin{eqnarray}
 m_h \sim 4 \pi v \sim 3~{\rm TeV},
\end{eqnarray}
which is quite different from the situation in the electroweak
sector. On the other hand, the MSSM provides a relation
\begin{eqnarray}
 m_h \sim m_Z \ll 4 \pi v.
\end{eqnarray}
Although this turns out to be too small, SUSY provides an excellent
starting point to explain 125~GeV; $m_Z \sim 91$~GeV rather than 3~TeV.

\section{Lessons from QCD}

The QCD is a natural place to look for a hint for the theory of
electroweak symmetry breaking. Other than the electroweak, that is the
only known example of the spontaneous symmetry breaking, actually
happening in the vacuum.
Indeed, the patterns of the symmetry breaking in the chiral symmetry
breaking and the electroweak symmetry breaking are identical; $SU(2)_L
\times U(1)_Y \to U(1)_{\rm EM}$. The only big difference is the mass of
$0^+$ resonances, {\it i.e.,} the Higgs boson mass relative to the order
parameter $v$ or $f_\pi$.

When we look up the table in the PDG~\cite{Beringer:1900zz}, we actually
find relatively light scalar mesons:
\begin{eqnarray}
f_0(600),\ \ \ f_0(980),
\end{eqnarray}
in which $f_0(600)$ is a pretty broad resonance. These are the
candidates of the Higgs boson counterparts in QCD. There are also light
and narrow vector resonances
\begin{eqnarray}
 \rho(770),\ \ \ \omega(782).
\end{eqnarray}
Of course, there are pions, $\pi^0$ and $\pi^\pm$ associated with the
chiral symmetry breaking, corresponding to the longitudinal modes of
$Z^0$ and $W^\pm$ in the electroweak symmetry breaking.

One can construct an effective theory to describe the interactions among
these light hadrons as the Higgs model~\cite{Kitano:2011zk,
Kitano:2012zz}, which is the (linearly realized) Hidden Local
Symmetry~\cite{Bando:1984ej}. The model is based on a $U(2)$ gauge
theory where we introduce Higgs fields:
\begin{eqnarray}
 H_L^{iA},\ \ \ H_R^{iA},\ \ \ (i,A=1,2).
\end{eqnarray}
Both are $2 \times 2$ matrices. One of the indices is for $U(2)$ gauge
group and the other is for flavor symmetry, $SU(2)_L \times
SU(2)_R$. The $H_L$ and $H_R$ fields transform under $SU(2)_L$ and
$SU(2)_R$, respectively. By requiring the flavor symmetry in the
Lagrangian and setting up the potential so that the minimum is at
\begin{eqnarray}
 H_L = H_R \propto {\bf 1},
\end{eqnarray}
the chiral symmetry is spontaneously broken down to a diagonal group,
and provides massless pions. The radial direction corresponds to the
Higgs boson, $f_0(980)$ (or $f_0(600)$), and the $U(2)$ gauge fields
obtain masses which correspond to $\rho(770)$ and $\omega(782)$.

This simple linear sigma model contains a pretty interesting object, a
string associated with the gauge symmetry breaking. By setting the
Lagrangian parameters so that the masses and the couplings of hadrons
are reproduced, one can estimate the tension of the string as
\cite{Kitano:2012zz}:
\begin{eqnarray}
 \sigma \sim (400~{\rm MeV})^2.
\label{eq:tension}
\end{eqnarray}
In QCD, there is a natural object to be identified with this string. It
is the QCD string which stretches between color sources to cause
confinement. The tension of the QCD string is estimated by the lattice
simulations and also from the quarkonium spectrum, and both give
approximately $\sigma \sim (430~{\rm MeV})^2$, which is very consistent
with Eq.~\eqref{eq:tension}.

If the string in the linear sigma model corresponds to the QCD string,
there appears an interesting interpretation of the Hidden Local
Symmetry. The QCD string is supposed to carry the color flux whereas the
string in the sigma model is carrying the magnetic flux of the $\rho$
and $\omega$ mesons.
The identification then says that the $\rho$ and $\omega$ mesons are the
magnetic gauge bosons in QCD. Since the Higgs fields $H_L$ and $H_R$ are
charged under $U(2)$, they are (non-abelian) magnetic monopoles.
The Higgs phenomena of the magnetic gauge theory, that is also
describing the chiral symmetry breaking, is now identified as the color
confinement in QCD.

There is a theoretical support for this interpretation.
It has been demonstrated in Ref.~\cite{Kitano:2011zk}, that by making a
deformation of a SUSY QCD model, one can see that the Hidden Local
Symmetry appears as the magnetic theory via the Seiberg duality. 
See also earlier discussion~\cite{Seiberg:1995ac, Harada:1999zj,
Komargodski:2010mc} on the interpretation of the $\rho$ meson as the
magnetic gauge boson.

The Seiberg dual picture of QCD is a $U(2)$ gauge theory with five
flavors, that is in fact identical to the model discussed in
Section~\ref{sec:model} for the model of dynamical electroweak symmetry
breaking.
The Higgs fields introduced above, $H_L$ and $H_R$ correspond to $f$ and
$\bar f$, respectively, in the model of Section~\ref{sec:model} which
arise as dual quarks transforming under both the magnetic gauge group
and the flavor group.
By adding soft SUSY breaking terms, the dual quarks obtain VEVs and the
magnetic gauge group is Higgsed while the chiral symmetry breaking
happens.
This Higgs mechanism in the magnetic picture corresponds to the
confinement.

Interestingly, a model of the low energy QCD can be regarded as the
magnetic picture, and the magnetic picture has the same structure as the
model of partially composite Higgs for the electroweak symmetry
breaking.
From this view, physics behind the electroweak symmetry breaking may
actually be very similar to (or the same as) QCD, and the quantitative
difference, the mass of the Higgs boson, may originate simply from
whether SUSY is a good (approximate) symmetry at the dynamical scale.

\section{Predictions}

There are two kinds of predictions in the framework we discussed. One is
the SUSY spectrum in Eq.~\eqref{eq:susy}. Since these are the estimates
from (probably unavoidable) finite quantum corrections, the naturalness
suggests that the estimates roughly give the upper bounds. The light
stop should be observed at the LHC soon.

Another interesting prediction is the presence of the relatively light
resonances from the strong sector. As we have discussed, the light Higgs
fields actually suggest a low dynamical scale such as $\Lambda \sim
600$~GeV, from the naive dimensional analysis.
Since the dynamics should have $SU(2)_L \times U(1)_Y$ as a part of the
global symmetry, one expects vector resonances, $W^\prime$ and
$Z^\prime$, associated with the current operators of the global
symmetry. These are the $\rho$ meson counterparts of QCD.
In the explicit model of Section~\ref{sec:model}, they corresponds to
$U(2)$ gauge bosons. Finding such resonances will be very important for
the understanding of the physics of electroweak symmetry breaking.

\section{Summary}

The limits on SUSY particles from LHC and the 125~GeV new particle may
be suggesting a very interesting possibility, a dynamical electroweak
symmetry breaking much below $4 \pi v \sim 3$~TeV.
In such a case, we will see many resonances and also superparticles at
the LHC, opening up a new era of particle physics to look for the theory
behind the electroweak symmetry breaking, just as it has been done to
look for QCD since the days of the discovery of the pions.

\section*{Acknowledgments}

I would like to thank organizers of HCP2012.  This work is supported in
part by the Grant-in-Aid for Scientific Research 23740165 of JSPS.

\end{document}